\documentclass[preprint,3p]{elsarticle}
\usepackage{amssymb}
\usepackage{amsthm}
\usepackage[mathlines]{lineno}
\usepackage{graphicx}
\usepackage{units}
\usepackage{url}
\usepackage{amsmath}
\usepackage{amsfonts}
\usepackage{bm}
\usepackage{textcomp}
\usepackage{subfigure}
\usepackage{multicol}
\usepackage{verbatim}
\usepackage{rotating}
\usepackage[colorlinks,linkcolor=red,citecolor=red]{hyperref}
\usepackage{float}
\usepackage{epsfig}
\usepackage{dcolumn}
\usepackage{bm}
\usepackage{color}
\usepackage{pstricks}
\usepackage{pst-node}
\usepackage{times}
\usepackage{indentfirst}
\usepackage[english]{babel}
\addto{\captionsenglish}{%

}
\journal{Physics Letters B}


\begin{document}
\begin{frontmatter}
\title{{\bf
\boldmath Measurement of the Absolute Branching Fraction for $\Lambda_c^+\rightarrow \Lambda \mu^+\nu_{\mu}$}}

\author{
\begin{small}
\begin{center}
      M.~Ablikim$^{1}$, M.~N.~Achasov$^{9,e}$, S.~Ahmed$^{14}$,
      X.~C.~Ai$^{1}$, O.~Albayrak$^{5}$, M.~Albrecht$^{4}$,
      D.~J.~Ambrose$^{44}$, A.~Amoroso$^{49A,49C}$, F.~F.~An$^{1}$,
      Q.~An$^{46,a}$
% ...
      \\
      \vspace{0.2cm}
      (BESIII Collaboration)\\
      \vspace{0.2cm} {\it
        $^{1}$ Institute of High Energy Physics, Beijing 100049, People's Republic of China\\
% ...
$^{4}$ Bochum Ruhr-University, D-44780 Bochum, Germany\\
        $^{5}$ Carnegie Mellon University, Pittsburgh, Pennsylvania 15213, USA\\
% ...
$^{9}$ G.I. Budker Institute of Nuclear Physics SB RAS (BINP), Novosibirsk 630090, Russia\\
% ...
$^{14}$ Helmholtz Institute Mainz, Johann-Joachim-Becher-Weg 45, D-55099 Mainz, Germany\\
% ...
$^{44}$ University of Rochester, Rochester, New York 14627, USA\\
% ...
$^{46}$ University of Science and Technology of China, Hefei 230026, People's Republic of China\\
% ...
$^{49}$ (A)University of Turin, I-10125, Turin, Italy; (B)University of Eastern Piedmont, I-15121, Alessandria, Italy; (C)INFN, I-10125, Turin, Italy\\
% ...
$^{a}$ Also at State Key Laboratory of Particle Detection and Electronics, Beijing 100049, Hefei 230026, People's Republic of China\\
% ...
$^{e}$ Also at the Novosibirsk State University, Novosibirsk, 630090, Russia\\
% ...
}\end{center}
\vspace{0.4cm}
\end{small}
}

\begin{abstract}
We report the first measurement of the absolute branching fraction for
$\Lambda^+_{c}\rightarrow \Lambda \mu^+\nu_{\mu}$. This measurement
is based on a sample of $e^+e^-$ annihilation data at a center-of-mass energy of
$\sqrt{s}=4.6$ GeV collected with the BESIII detector at the BEPCII storage rings. The sample corresponds to an integrated luminosity of 567 pb$^{-1}$.
The branching
fraction is determined to be $\mathcal B({\Lambda^+_c\rightarrow
\Lambda \mu^+\nu_{\mu}})=(3.49\pm0.46({\rm stat})\pm0.27({\rm
syst}))\%$. In addition, we  calculate the ratio
$\mathcal{B}(\Lambda^+_c\rightarrow \Lambda
\mu^+\nu_{\mu})/\mathcal{B}(\Lambda^+_c\rightarrow \Lambda e^+\nu_{e})$
 to be $0.96\pm0.16({\rm stat})\pm0.04({\rm syst})$.
\\
\\
\text{Keywords:~~$\Lambda_c^+$, semi-leptonic decay, absolute branching fraction, BESIII}
\end{abstract}

\end{frontmatter}

\begin{multicols}{2}

\section{Introduction}
Semileptonic (SL) decays of the lightest charmed baryon,
$\Lambda_c^+$, provide a stringent test for non-perturbative aspects
of the strong interaction theory. The $\Lambda^+_c\rightarrow
\Lambda \ell^+\nu_{\ell}$ ($\ell$ denotes lepton) decay is dominated by the Cabibbo-favored
transition $c\rightarrow s \ell^+\nu_\ell$, which occurs independently of the spin-zero and isospin-zero spectator $ud$
diquark, to good
approximation. This leads to a simpler theoretical description and greater
predictive power in the non-perturbative models
than in the case for charmed mesons~\cite{Richman:1995wm}. Predictions of the branching fraction
(BF) $\mathcal{B}(\Lambda_c^+\rightarrow \Lambda \ell^+\nu_\ell)$ in different
theoretical models vary over a wide range from 1.4\% to
9.2\%~\cite{prd40_2955,prd40_2944,zpc51_607,zpc52_149,prd43_2939,prd45_3266,prd53_1457,plb_431_173,prd60_034009,prc72_032005,prd80_074011,prd93_034008},
depending on the choice of $\Lambda_c^+$ wave function
model and the treatment of decay dynamics. In 2015, BESIII  measured the absolute BF for $\Lambda^+_c\rightarrow
\Lambda e^+\nu_e$ to be $\mathcal B({\Lambda^+_c\rightarrow
\Lambda e^+\nu_e})=(3.63\pm0.38\pm0.20)\%$~\cite{bes3lamev}, which disfavors
the predictions in
Refs.~\cite{prd40_2955,prd40_2944,zpc52_149,prd43_2939,prd45_3266} at the 95\% confidence level.
It is desirable to test these theoretical predictions by measuring $\mathcal{B}(\Lambda^+_c\rightarrow
\Lambda \mu^+\nu_{\mu})$. In addition, lepton universality can be tested by comparing the BFs between the electronic and muonic modes.

In this paper, we report the first absolute measurement
of $\mathcal{B}(\Lambda_c^+\rightarrow \Lambda \mu^+\nu_{\mu})$ by
analyzing a data sample with an integrated luminosity of $567$ pb$^{-1}$~\cite{lum} accumulated at a center-of-mass (c.m.) energy of
$\sqrt{s} = 4.6$ GeV with the BESIII detector at the BEPCII
collider, which is the largest $e^+e^-$ collision sample near the
$\Lambda^+_c\bar{\Lambda}^-_c$ mass threshold. At this energy, the $\Lambda^+_c$ is
produced in company with one $\bar{\Lambda}^-_c$ baryon only, and no other hadrons are kinematically allowed.
Hence, $\mathcal{B}(\Lambda^+_{c}\rightarrow \Lambda \mu^+\nu_{\mu})$ can be
accessed by measuring the relative probability of finding the SL decay
when the $\bar{\Lambda}^-_c$ is detected in a number of prolific decay
channels. This will provide a straightforward and direct BF
measurement without requiring knowledge of the total number of
$\Lambda^+_c\bar{\Lambda}^-_c$ pairs produced.
In the following, charge conjugated modes are always implied, unless explicitly mentioned.

\section{BESIII Detector and Monte Carlo Simulation}
The BESIII~\cite{Ablikim:2009aa} detector is a cylindrical detector with
a solid-angle coverage of 93\% of $4\pi$ that operates at the
BEPCII collider. It consists of a Helium-gas based main drift chamber (MDC), a plastic
scintillator time-of-flight (TOF) system, a CsI~(Tl) electromagnetic
calorimeter (EMC), a superconducting solenoid providing a 1.0\,T
magnetic field and a muon counter. The  charged particle momentum
resolution is 0.5\% at a transverse momentum of 1~GeV/$c$. The
photon energy resolution in the EMC is 2.5\% in the barrel and 5.0\% in the end-caps at 1\,GeV.
More details about the design and performance of the detector are given in
Ref.~\cite{Ablikim:2009aa}.

A GEANT4-based~\cite{geant4} Monte Carlo (MC) simulation package,
which includes the geometric description of the detector and the
detector response, is used to determine the detection efficiency and
to estimate the potential backgrounds. Signal MC samples of a
$\Lambda^+_c$ baryon decaying only to $\Lambda \mu^+\nu_{\mu}$ together with a
$\bar{\Lambda}^-_c$ decaying to specified modes are generated
with the KKMC~\cite{kkmc} and
EVTGEN~\cite{nima462_152}, taking into account the initial state radiation (ISR)~\cite{SJNP41_466} and the final state radiation (FSR)~\cite{plb303_163} effects.
For the simulation of the process $\Lambda_c^+\rightarrow \Lambda
\mu^+\nu_{\mu}$, we use the form factor obtained using Heavy Quark
Effective Theory and QCD sum rules in Ref.~\cite{prd60_034009}.
To study backgrounds, inclusive MC samples are simulated which consist of $\Lambda_c^+\bar{\Lambda}_c^-$
events, $D_{(s)}^{*}\bar{D}_{(s)}^{(*)}+X$ production, ISR return to the charmonium(-like)
$\psi$ states at lower masses, and QED processes.
The decay modes with known BFs of the $\Lambda_c$, $\psi$ and $D_{(s)}$ particles taken from Particle Data Group (PDG)~\cite{pdg2014} are simulated with EVTGEN, while the remaining unknown decays are generated with LUNDCHARM~\cite{lundcharm}.

\section{Analysis}

Following the similar technique of the single tag (ST) and double tag (DT) in Ref.~\cite{bes3lamev},
we reconstruct the $\bar{\Lambda}^-_c$ baryons in eleven hadronic
decay modes as listed in the first column of Table~\ref{tab:deltaE_1}.
The intermediate particles $K^0_S$, $\bar{\Lambda}$, $\bar{\Sigma}^0$,
$\bar{\Sigma}^-$ and $\pi^0$ are reconstructed through their decays
$K^0_S\rightarrow \pi^+\pi^-$, $\bar{\Lambda}\rightarrow
\bar{p}\pi^+$, $\bar{\Sigma}^0\rightarrow \gamma\bar{\Lambda}$ with
$\bar{\Lambda}\rightarrow \bar{p}\pi^+$, $\bar{\Sigma}^-\rightarrow
\bar{p}\pi^0$ and $\pi^0\rightarrow \gamma\gamma$, respectively.
The detailed selection criteria for charged and neutral tracks, $\pi^0$, $K^0_S$ and $\bar{\Lambda}$ candidates
used in the reconstruction of tags are described in Ref.~\cite{bes3lamev}.

In this analysis, the ST $\bar{\Lambda}^-_c$ signals are identified using the beam energy
constrained mass, $M_{\rm BC}=\sqrt{E^2_{\rm
beam}/c^4-|\vec{p}_{\bar{\Lambda}^-_c}|^2/c^2}$, where $E_{\rm
beam}$ is the beam energy and
$\vec{p}_{\bar{\Lambda}^-_c}$ is the momentum of the
$\bar{\Lambda}^-_c$ candidate. To improve the signal purity, the
energy difference $\Delta E=E_{\rm beam}-E_{\bar{\Lambda}^-_c}$ for
each candidate is required to be within
$\pm3\sigma_{\Delta E}$ around the $\Delta E$ peak, where
$\sigma_{\Delta E}$ is the $\Delta E$ resolution and
$E_{\bar{\Lambda}^-_c}$ is the reconstructed $\bar{\Lambda}^-_c$
energy. Table~\ref{tab:deltaE_1} shows the mode dependent
$\Delta E$ requirements and the ST yields in the $M_{\rm BC}$ signal
region $(2.280, 2.296)$~GeV/$c^2$,
which are obtained by a fit to the $M_{\rm BC}$ distributions.
The detailed process to extract the ST signal yields is described in Ref.~\cite{bes3lamev}.
The total ST yield summed over all 11 modes is
$N^{\rm tot}_{\bar{\Lambda}^-_c}=14415\pm159$, where the uncertainty is statistical only.

\begin{table*}[htp]
\centering
\caption{\label{tab:deltaE_1}$\Delta E$ requirements and  ST yields
$N_{\bar{\Lambda}_c^-}$ in data, in which the uncertainties are statistical only.}
\begin{tabular}{llc} \hline \hline Mode~&~~~$\Delta E$~(GeV)~~~&$N_{\bar{\Lambda}_c^-}$ \\
\hline
 $\bar{p} K^0_S$                & [$-$0.025, 0.028] &   $1066\pm33$  \\
 $\bar{p} K^+\pi^-$             & [$-$0.019, 0.023] &   $5692\pm88$  \\
 $\bar{p}K^0_S\pi^0$            & [$-$0.035, 0.049] &  ~~$593\pm41$  \\
 $\bar{p} K^+\pi^-\pi^0$        & [$-$0.044, 0.052] &   $1547\pm61$  \\
 $\bar{p} K^0_S\pi^+\pi^-$      & [$-$0.029, 0.032] &  ~~$516\pm34$  \\
 $\bar{\Lambda}\pi^-$           & [$-$0.033, 0.035] &  ~~$593\pm25$  \\
 $\bar{\Lambda}\pi^-\pi^0$      & [$-$0.037, 0.052] &   $1864\pm56$  \\
 $\bar{\Lambda}\pi^-\pi^+\pi^-$ & [$-$0.028, 0.030] &  ~~$674\pm36$  \\
 $\bar{\Sigma}^0\pi^-$          & [$-$0.029, 0.032] &  ~~$532\pm30$  \\
 $\bar{\Sigma}^-\pi^0$          & [$-$0.038, 0.062] &  ~~$329\pm28$  \\
 $\bar{\Sigma}^-\pi^+\pi^-$     & [$-$0.049, 0.054] &   $1009\pm57$  \\
\hline \hline
\end{tabular}
\end{table*}
Candidate events for $\Lambda^+_c\rightarrow \Lambda \mu^+\nu_{\mu}$ are
selected from the remaining tracks recoiling against the ST
$\bar{\Lambda}^-_c$ candidates. The $\Lambda$ candidate is formed from a $p\pi^-$ combination that is constrained by a common vertex fit to have a positive decay length $L$.
If multiple $\Lambda$ candidates are formed, the one with the largest $L/\sigma_L$ is retained, where $\sigma_L$ is the resolution of the measured $L$.
Particle identification (PID) is performed using probabilities derived from the specific energy loss $dE/dx$ measured by the MDC, the time of flight measured by the TOF, and energy measured by the EMC; a $\mu$ candidate is required to satisfy
$\mathcal{L}'_{\mu} > 0.001$, $\mathcal{L}'_{\mu} > \mathcal{L}'_{e}$ and
$\mathcal{L}'_{\mu} > \mathcal{L}'_{K}$, where $\mathcal{L}'_{\mu}$, $\mathcal{L}'_{e}$, and $\mathcal{L}'_{K}$ are the probabilities for a muon, electron, and kaon, respectively.

Studies on the inclusive MC samples show that the backgrounds are dominated by $\Lambda_c^+\rightarrow
\Lambda\pi^+$, $\Sigma^0\pi^+$ and $\Lambda\pi^+\pi^0$.
Backgrounds from $\Lambda_c^+\rightarrow\Lambda\pi^+$ and $\Lambda_c^+\rightarrow \Sigma^0\pi^+$ are rejected by
requiring the $\Lambda\mu^+$ invariant mass $M_{\Lambda\mu^+}$ is less than $2.12$ GeV/c$^2$. The background from
$\Lambda_c^+\rightarrow \Lambda\pi^+\pi^0$ is suppressed by requiring the largest energy of
any unused photons $E_{\gamma \rm{max}}$
be less than 0.25 GeV and the deposited
energy for the muon candidate in the EMC be less than 0.30 GeV.

Since the neutrino is not detected, we employ the kinematic variable
$U_{\rm miss} \equiv E_{\rm miss}-|\vec{p}_{\rm miss}\: c|$ to identify the neutrino signal, where $E_{\rm
miss}$ and $\vec{p}_{\rm miss}$ are the missing energy and momentum
carried by the neutrino, respectively. They are calculated as
$E_{\rm miss}=E_{\rm beam}-E_{\Lambda}-E_{\mu^+}$ and $\vec{p}_{\rm
miss}=\vec{p}_{\Lambda_c^+}-\vec{p}_{\Lambda}-\vec{p}_{\mu^+}$, where
$\vec{p}_{\Lambda_c^+}$ is the momentum of the $\Lambda_c^+$ baryon, $E_{\rm
\Lambda}$($\vec{p}_{\Lambda}$) and $E_{\mu^+}$($\vec{p}_{\mu^+}$) are
the energies (momenta) of the $\Lambda$ and $\mu^+$,
respectively. Here, the momentum $\vec{p}_{\Lambda_c^+}$ is given by
$\vec{p}_{\Lambda_c^+}=-\hat{p}_{\rm tag}\sqrt{E_{\rm
beam}^2/c^2-m^2_{\bar{\Lambda}^-_c}}$, where $\hat{p}_{\rm tag}$ is the
momentum direction of the ST $\bar{\Lambda}^-_c$ and
$m_{\bar{\Lambda}^-_c}$ is the nominal $\bar{\Lambda}^-_c$
mass~\cite{pdg2014}. For the signal events, the $U_{\rm miss}$ distribution
is expected to peak at zero.

\begin{figure*}[htp]
  \centering
   \begin{minipage}[t]{7.2cm}
   \includegraphics[height=4.7cm,width=6.4cm]{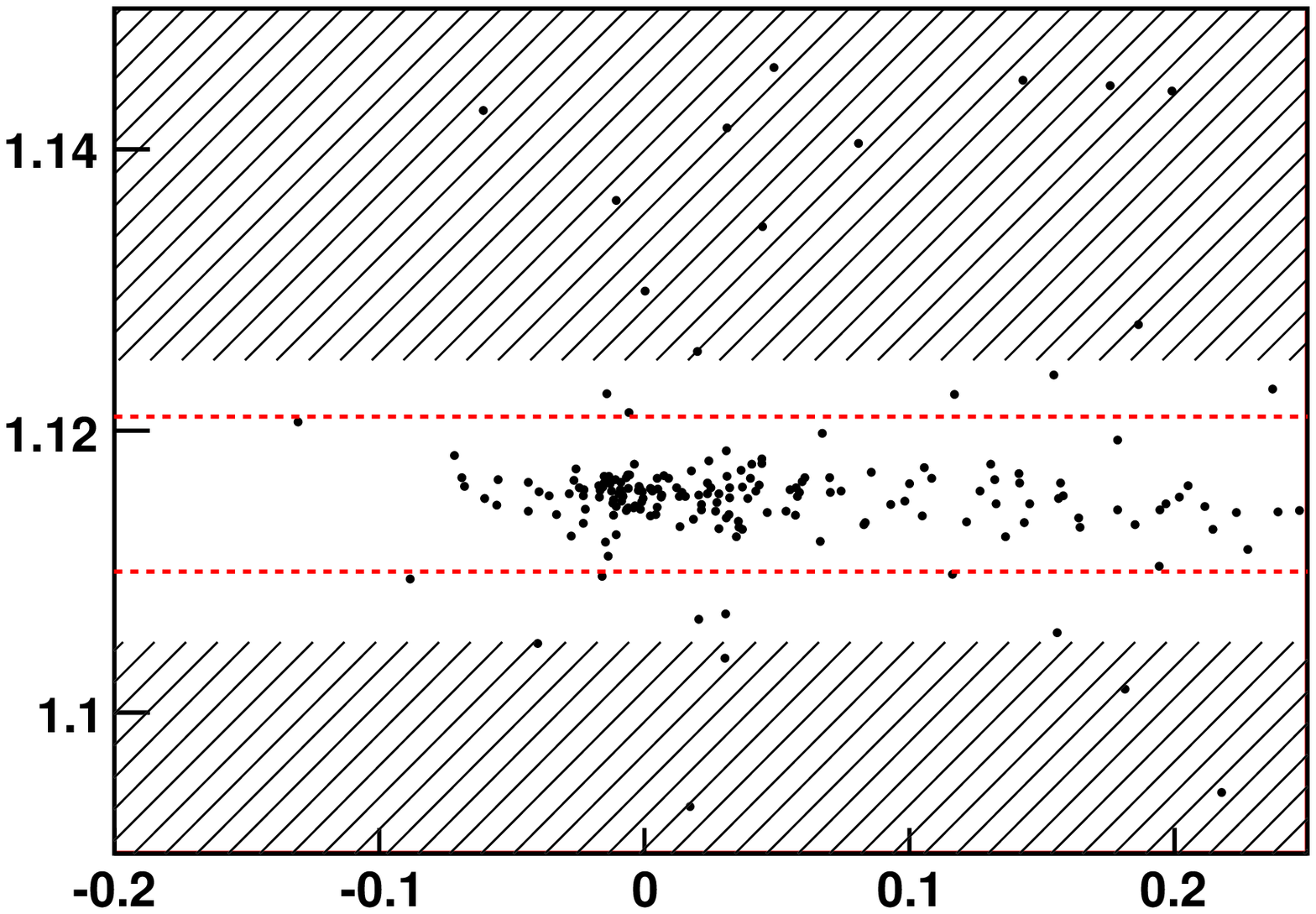}
   \put(-47, 95){\bf \normalsize (a)}
   \put(-190,38){\rotatebox{90}{\normalsize $M_{p\pi^-}$ (GeV$/c^2$)}}
   \put(-110,-4){\normalsize $U_{\rm miss}$ (GeV) }
   \end{minipage}
   \begin{minipage}[t]{7.2cm}
   \includegraphics[height=4.7cm,width=6.4cm]{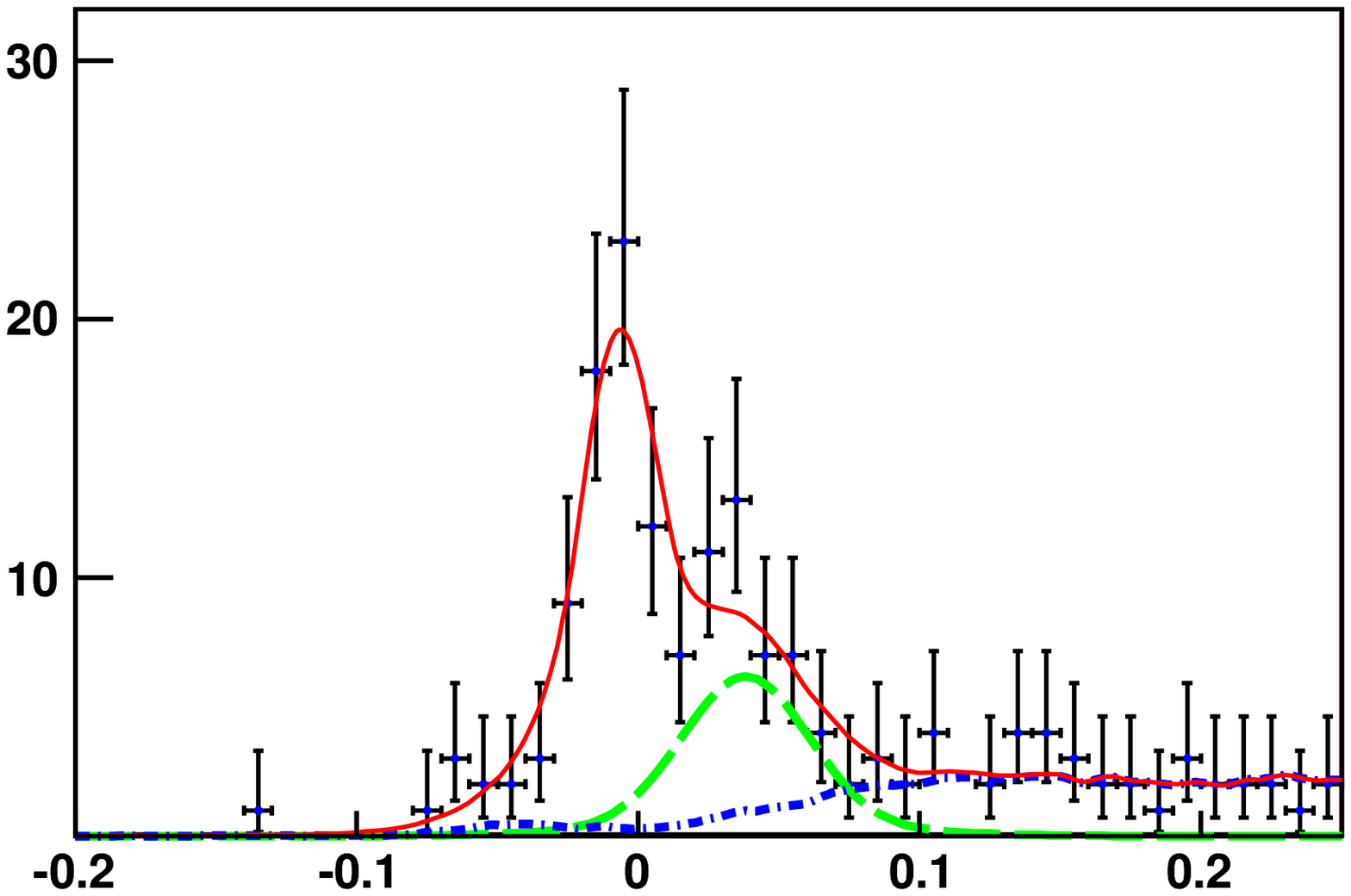}
   \put(-47, 95){\bf \normalsize (b)}
   \put(-190,38){\rotatebox{90}{\normalsize Events/0.010 GeV}}
   \put(-110,-4){\normalsize $U_{\rm miss}$ (GeV) }
   \end{minipage}
   \caption{ (a) Distribution of $M_{p\pi^-}$ versus $U_{\rm miss}$ for the $\Lambda_c^+\rightarrow \Lambda \mu^+\nu_{\mu}$
   candidates in data. The area between the dashed lines denotes the $\Lambda$ signal region and the hatched areas indicate the $\Lambda$ sideband regions. (b) Fit to the $U_{\rm miss}$ distribution within the $\Lambda$ signal region. Data are shown as the dots with error bars. The
   long-dashed curve (green) shows the $\Lambda_c^+\rightarrow\Lambda\pi^+\pi^0$ background while
   the dot-dashed curve (blue) shows other $\Lambda_c^+$ decay backgrounds. The thick line (red) shows the total fit.}
\label{fig:umiss_data}
\end{figure*}

The distribution of the $p\pi^-$ invariant mass $M_{p\pi^-}$ versus $U_{\rm miss}$ for the $\Lambda^+_c\to
\Lambda\mu^+\nu_{\mu}$ candidates in data is
shown in Fig.~\ref{fig:umiss_data} (a), where a cluster around the signal region is evident.
After requiring $M_{p\pi^-}$ to be within the $\Lambda$ signal
region, the projection of $U_{\rm miss}$ is shown in Fig.~\ref{fig:umiss_data}(b).
Two bumps, which correspond to the signal peak (left side) and background $\Lambda_c^+\rightarrow \Lambda\pi^+\pi^0$ (right side), are visible.
According to MC simulations, the survival
rate of the background process $\Lambda_c^+\rightarrow \Lambda\pi^+\pi^0$ is estimated to be $\eta_{\Lambda\pi^+\pi^0}=(3.67\pm0.05)\%$, where the BFs
for $\Lambda\rightarrow p\pi^-$ and $\pi^0\rightarrow \gamma\gamma$ are included.
Thus, the number of the $\Lambda_c^+\rightarrow \Lambda\pi^+\pi^0$
background events can be
estimated by
\begin{equation}
N^{\rm bkg}_{\Lambda \pi^+\pi^0}=N^{\rm
tot}_{\bar{\Lambda}_c^-}\cdot\mathcal{B}(\Lambda^+_c\rightarrow \Lambda \pi^+\pi^0)\cdot\eta_{\Lambda \pi^+\pi^0}.
\label{eq:equa_bkg}
\end{equation}
Inserting the values of $N^{\rm tot}_{\bar{\Lambda}^-_c}$, $\eta_{\Lambda \pi^+\pi^0}$ and
$\mathcal{B}(\Lambda^+_c\rightarrow \Lambda \pi^+\pi^0)=(7.01\pm0.42)\%$~\cite{1511.08380} in
Eq.~(\ref{eq:equa_bkg}), we obtain $N^{\rm bkg}_{\Lambda \pi^+\pi^0}=37.1\pm2.3$, where the uncertainties from
$N^{\rm tot}_{\bar{\Lambda}^-_c}$, $\eta_{\Lambda \pi^+\pi^0}$ and
$\mathcal{B}(\Lambda^+_c\rightarrow \Lambda \pi^+\pi^0)$ are included.

We apply a fit to the $U_{\rm miss}$ distribution to obtain the signal yields. The $\Lambda_c^+\rightarrow \Lambda\mu^+\nu_{\mu}$ signal shape is described with a function $f$, which consists of a Gaussian function to model the
core of the $U_{\rm miss}$ distribution and two power law tails to
account for the effects of ISR and FSR in the form~\cite{prd79_052010} of
\begin{equation}
f(U_{\rm miss})=\left\{
\begin{array}{rcl}
p_1(\frac{n_1}{\alpha_1}-\alpha_1+t)^{-n_1},      &      & t>\alpha_1\\
e^{-t^2/2},~~~~~~~~~~~~~~~~     &      & -\alpha_2<t<\alpha_1 . \\
p_2(\frac{n_2}{\alpha_2}-\alpha_2-t)^{-n_2},       &      &
t<-\alpha_2
\end{array} \right.
\label{eq:fun_umiss}
\end{equation}
Here, $t\equiv(U_{\rm miss}-U_{\rm mean})/\sigma_{U_{\rm miss}}$, $U_{\rm
mean}$ and $\sigma_{U_{\rm miss}}$ are the mean value and resolution
of the Gaussian function, respectively,
$p_1\equiv(n_1/\alpha_1)^{n_1}e^{-\alpha^2_1/2}$ and
$p_2\equiv(n_2/\alpha_2)^{n_2}e^{-\alpha^2_2/2}$. The parameters
$\alpha_1$, $\alpha_2$, $n_1$ and $n_2$ are fixed to the values
obtained by studying the signal MC simulations.
For backgrounds, a double Gaussian function with parameters fixed according to MC simulations is used to describe
the $\Lambda_c^+\rightarrow \Lambda\pi^+\pi^0$ peaking background and
a MC-derived shape is used to describe other combinatorial backgrounds.
In the fit, we fix the number of the
$\Lambda^+_c\rightarrow \Lambda \pi^+\pi^0$ background events to be estimated $N^{\rm bkg}_{\Lambda \pi^+\pi^0}$ as described above.
From the fit, we obtain the
number of events of $\Lambda_c^+\rightarrow \Lambda \mu^+\nu_{\mu}$
to be $N^{\rm obs}_{\Lambda\mu^+\nu_{\mu}}=78.7\pm10.5$, where the uncertainty is statistical only.
A fit with unconstrained $N^{\rm bkg}_{\Lambda \pi^+\pi^0}$ gives $77.1\pm11.4$ events of signal, which is in good agreement with the estimation when  $N^{\rm bkg}_{\Lambda \pi^+\pi^0}$ is fixed. Based on the data in $\Lambda$ sidebands in Fig.~\ref{fig:umiss_data}(a),
the background events from the non-$\Lambda$ SL decays is found to be negligible.

The absolute BF for $\Lambda_c^+\rightarrow \Lambda \mu^+\nu_{\mu}$ is determined with
\begin{equation}
\mathcal{B}(\Lambda_c^+\rightarrow \Lambda \mu^+\nu_{\mu})=\frac{N^{\rm obs}_{\Lambda\mu^+\nu_{\mu}}}{N^{\rm tot}_{\bar{\Lambda}_c^-}\cdot\varepsilon_{\Lambda \mu^+\nu_{\mu}}\cdot\mathcal{B}(\Lambda\rightarrow p\pi^-)},
\label{eq:branch}
\end{equation}
where $\varepsilon_{\Lambda \mu^+\nu_{\mu}}$ is the detection efficiency for
the $\Lambda_c^+\rightarrow \Lambda \mu^+\nu_{\mu}$ decay, which does not
include the BF for $\Lambda\rightarrow p\pi^-$.
For each ST mode $i$, the efficiency $\varepsilon^i_{\Lambda \mu^+\nu_{\mu}}$ is
obtained by dividing the DT efficiency $\varepsilon^i _{{\rm tag}, \Lambda \mu^+\nu_{\mu}}$ by
the ST efficiency $\varepsilon^i _{\rm tag}$.
After weighting $\varepsilon^i_{\Lambda \mu^+\nu_{\mu}}$ with the ST yields in data for each ST mode $i$,
we determine the overall average efficiency $\varepsilon_{\Lambda \mu^+\nu_{\mu}}=(24.5\pm0.2)\%$.
By inserting the values of $N^{\rm obs}_{\Lambda\mu^+\nu_{\mu}}$, $N^{\rm tot}_{\bar{\Lambda}^-_c}$, $\varepsilon_{\Lambda\mu^+\nu_{\mu}}$ and
$\mathcal{B}(\Lambda\rightarrow p\pi^-)$~\cite{pdg2014} in
Eq.~(\ref{eq:branch}),
we obtain
$\mathcal B(\Lambda_c^+\rightarrow \Lambda\mu^+\nu_{\mu})=(3.49\pm0.46\pm0.27)\%$, where the first uncertainty is
statistical, and the second uncertainty is  systematic  as described below.

With the DT technique, the uncertainties on the BF measurement are insensitive to those originating from the ST side.
The systematic uncertainties for measuring
$\mathcal{B}(\Lambda_c^+\rightarrow \Lambda\mu^+\nu_{\mu})$ mainly arise from
the uncertainties related to the tracking and PID of the muon candidate, $\Lambda$ reconstruction,
$U_{\rm miss}$ fit, peaking background subtraction, $E_{\gamma{\rm max}}$ and $M_{\Lambda\mu^+}$ requirements, and signal MC modeling.
Throughout this paragraph, the systematic uncertainties quoted
are relative uncertainties.
The uncertainties of the $\mu^+$ tracking and PID are
determined to be 1.0\% and 2.0\%, respectively, by studying a control sample of
$e^+e^-\rightarrow (\gamma)\mu^+\mu^-$ events.
The uncertainty of the $\Lambda$ reconstruction is determined to be 2.5\% by
studying a control sample of $\chi_{cJ}\rightarrow
\Lambda\bar{\Lambda}\pi^+\pi^-$ decays.
The uncertainty of $U_{\rm miss}$ fit is estimated to be 1.5\% obtained by varying the fitting range and examining the fluctuation of the non-peaking background shape. The uncertainty due to peaking background $\Lambda_c^+\rightarrow\Lambda\pi^+\pi^0$ subtraction is estimated to be 2.5\% obtained by varying $N^{\rm bkg}_{\Lambda\pi^+\pi^0}$ equivalent variations of $\pm1\sigma$ of the quoted BFs, and smearing the MC-derived shape of $\Lambda_c^+\rightarrow \Lambda\pi^+\pi^0$ backgrounds with a Gaussian function to accommodate the resolution difference between the data and MC simulation.
The uncertainty in the $E_{\gamma \rm{max}}$ requirement
is estimated to be 2.6\% by using a control sample of $e^+e^-\rightarrow p\bar{p}\pi^+\pi^-$ events.
The uncertainty in the $M_{\Lambda\mu^+}$ requirement is estimated to be 2.0\% by comparing the obtained $\mathcal{B}(\Lambda^+_{c}\rightarrow \Lambda\mu^+\nu_{\mu})$ under the alternative requirements of $M_{\Lambda\mu^+}<2.07$~GeV/$c^2$ or $M_{\Lambda\mu^+}<2.17$~GeV/$c^2$ with the nominal value.
The uncertainty in the signal MC model is estimated to be 5.2\% by varying the parameterization of
the form factor function according to Refs.~\cite{prd60_034009,Hinson:2004pj}
and by taking into account the $q^2$ dependence observed in data.
In addition, there
are systematic uncertainties from the quoted $\mathcal{B}(\Lambda\rightarrow p\pi^-)$ (0.8\%), the $N^{\rm tot}_{\bar{\Lambda}_c^-}$
(1.0\%) evaluated by using alternative signal shapes in the fits to the
$M_{\rm BC}$ spectra~\cite{bes3lamev}, and MC statistics (0.8\%).
All these systematic uncertainties
are summarized in Table~\ref{tab:syst}, and the total systematic
uncertainty is evaluated to be 7.7\% by summing up all the individual contributions in
quadrature.

\begin{table*}[htp]
\caption{\label{tab:syst}\normalsize Summary of the relative systematic uncertainties for $\mathcal{B}(\Lambda^+_{c}\rightarrow \Lambda\mu^+\nu_{\mu})$.}
  \centering
\begin{tabular}
{lc} \hline \hline Source & Uncertainty \\ \hline
$\mu^{+}$ tracking & 1.0\% \\
$\mu^{+}$ PID & 2.0\%  \\
$\Lambda$ reconstruction & 2.5\% \\
$U_{\rm miss}$ fit  & 1.5\% \\
Peaking background $\Lambda_c^+\rightarrow\Lambda\pi^+\pi^0$ & 2.5\% \\
$E_{\gamma{\rm max}}$ requirement & 2.6\% \\
$M_{\Lambda\mu^+}$ requirement & 2.0\% \\
MC model   & 5.2\% \\
$\mathcal{B}(\Lambda\rightarrow p\pi^-)$ & 0.8\% \\
$N^{\rm tot}_{\Bar{\Lambda}_c^-}$ & 1.0\% \\
MC statistics  & 0.8\% \\
\hline Total  & 7.7\% \\
\hline \hline
\end{tabular}
\end{table*}

The ratio of branching fractions $\mathcal{B}(\Lambda^+_c\rightarrow \Lambda
\mu^+\nu_{\mu})/\mathcal{B}(\Lambda^+_c\rightarrow \Lambda e^+\nu_{e})$
is calculated with the measured $\mathcal{B}(\Lambda^+_c\rightarrow \Lambda \mu^+\nu_{\mu})$ in this
work and $\mathcal B({\Lambda^+_c\rightarrow
\Lambda e^+\nu_e})=(3.63\pm0.38({\rm stat})\pm0.20({\rm syst}))\%$ from BESIII~\cite{bes3lamev}. We determine $\mathcal{B}(\Lambda^+_c\rightarrow \Lambda
\mu^+\nu_{\mu})/\mathcal{B}(\Lambda^+_c\rightarrow \Lambda e^+\nu_{e})$
to be $0.96\pm0.16\pm0.04$, where the first uncertainty is
statistical and the second is systematic, in which the common systematic uncertainties from the tracking efficiency, the $\Lambda$ reconstruction,
the quoted BF for
$\Lambda\rightarrow p\pi^-$, the number of $\bar{\Lambda}_c^-$ tags $N^{\rm tot}_{\bar{\Lambda}_c^-}$ and the MC model cancel.

\section{Summary}

In summary, based on the $e^+e^-$ collision data corresponding to an integrated luminosity of 567~pb$^{-1}$ taken at
$\sqrt{s}=4.6$~GeV with the BESIII detector, we report the
first direct measurement of the absolute BF for $\Lambda^+_c\rightarrow \Lambda
\mu^+\nu_{\mu}$ to be $(3.49\pm0.46\pm0.27)\%$, where the first uncertainty is
 statistical  and the second is systematic.
The result is consistent with the value in PDG~\cite{pdg2014} within $2\sigma$ of uncertainty, but with improved precision.
This study helps to extend our understanding on the decay mechanism of the $\Lambda_c^+$ SL decay.
Based on this result and the previous BESIII work~\cite{bes3lamev}, we determine the ratio
$\mathcal{B}(\Lambda^+_c\rightarrow \Lambda
\mu^+\nu_{\mu})/\mathcal{B}(\Lambda^+_c\rightarrow \Lambda
e^+\nu_{e})=0.96\pm0.16\pm0.04$, which is compatible with unity.
As the theoretical predictions on
$\mathcal B(\Lambda^+_{c}\rightarrow \Lambda \ell^+\nu_\ell)$ vary
in a large range of $1.4\%$ to $9.2\%$~\cite{prd40_2955,prd40_2944,zpc51_607,zpc52_149,prd43_2939,prd45_3266,prd53_1457,plb_431_173,prd60_034009,prc72_032005,prd80_074011,prd93_034008},
the measured $\mathcal B(\Lambda^+_{c}\rightarrow \Lambda
\mu^+\nu_{\mu})$ in this work and $\mathcal B(\Lambda^+_{c}\rightarrow \Lambda e^+\nu_e)$ in Ref.~\cite{bes3lamev} provide stringent tests on these non-perturbative models.

\section{Acknowledgments}

The BESIII collaboration thanks the staff of BEPCII and the IHEP
computing center for their strong support. This work is supported in
part by National Key Basic Research Program of China under Contract
No.\ 2015CB856700; National Natural Science Foundation of China (NSFC)
under Contracts Nos.\ 11235005, 11235011, 11305090, 11322544, 11305180, 11335008, 11425524, 11505010; the
Chinese Academy of Sciences (CAS) Large-Scale Scientific Facility
Program; the CAS Center for Excellence in Particle Physics (CCEPP);
the Collaborative Innovation Center for Particles and Interactions
(CICPI); Joint Large-Scale Scientific Facility Funds of the NSFC and
CAS under Contracts Nos. U1232201, U1332201; CAS under Contracts
Nos. KJCX2-YW-N29, KJCX2-YW-N45; 100 Talents Program of CAS; National
1000 Talents Program of China; INPAC and Shanghai Key Laboratory for
Particle Physics and Cosmology; German Research Foundation DFG under
Contracts Nos. Collaborative Research Center CRC 1044, FOR 2359;
Istituto Nazionale di Fisica Nucleare, Italy; Joint Large-Scale
Scientific Facility Funds of the NSFC and CAS	under Contract
No.\ U1532257; Joint Large-Scale Scientific Facility Funds of the NSFC
and CAS under Contract No.\ U1532258; Koninklijke Nederlandse Akademie
van Wetenschappen (KNAW) under Contract No.\ 530-4CDP03; Ministry of
Development of Turkey under Contract No.\ DPT2006K-120470; NSFC under
Contract No.\ 11275266; The Swedish Resarch Council; U.S.\ Department
of Energy under Contracts Nos.\ DE-FG02-05ER41374, DE-SC-0010504,
DE-SC0012069, DESC0010118; U.S.\ National Science Foundation;
University of Groningen (RuG) and the Helmholtzzentrum fuer
Schwerionenforschung GmbH (GSI), Darmstadt; WCU Program of National
Research Foundation of Korea under Contract No.\ R32-2008-000-10155-0.
This paper is also supported by the Beijing municipal government under
Contract Nos.\ KM201610017009, 2015000020124G064.


\end{multicols}

\begin{thebibliography}{99}

\bibitem{Richman:1995wm}
  J.~D.~Richman and P.~R.~Burchat,  Rev.\ Mod.\ Phys.\  {\bf 67}, 893
  (1995); E. Eichten and B. Hill, Phys. Lett. B {\bf 234}, 511 (1990); M. Neubert, Phys. Rep. {\bf 245}, 259 (1994).

\bibitem{prd40_2955} R. P$\acute{\rm e}$rez-Marcial $et~al.$, Phys. Rev. D {\bf 40}, 2955 (1989).
\bibitem{prd40_2944} M. Avila-Aoki $et~al.$, Phys. Rev. D {\bf 40}, 2944 (1989).
\bibitem{zpc51_607} F. Hussain $et~al.$, Z. Phys. C {\bf 51}, 607 (1991).
\bibitem{zpc52_149} G. V. Efimov $et~al.$, Z. Phys. C {\bf 52}, 149 (1991).
\bibitem{prd43_2939} Robert Singleton, Phys. Rev. D {\bf 43}, 2939 (1991).
\bibitem{prd45_3266} A. Garcia and R. Huerta, Phys. Rev. D {\bf 45}, 3266 (1992).
\bibitem{prd53_1457} H. Y. Cheng and B. Tseng, Phys. Rev. D {\bf 53}, 1457 (1995).

\bibitem{plb_431_173} H. G. Dosch $et~al.$, Phys. Lett. B {\bf 431}, 173 (1998).
\bibitem{prd60_034009} R. S. Marques de Carvalho $et~al.$, Phys. Rev. D {\bf 60}, 034009 (1999).
\bibitem{prc72_032005} M. Pervin $et~al.$, Phys. Rev. C {\bf 72}, 035201 (2005).
\bibitem{prd80_074011} Y. L. Liu $et~al.$, Phys. Rev. D {\bf 80}, 074011 (2009).
\bibitem{prd93_034008} T. Gutsche $et~al.$, Phys. Rev. D {\bf 93}, 034008 (2016).

\bibitem{bes3lamev} M.~Ablikim $et~al.$ [BESIII Collaboration], Phys. Rev. Lett. {\bf 115}, 221805 (2015).



\bibitem{lum} M.~Ablikim $et~al.$ [BESIII Collaboration], Chin. Phys. C {\bf 39}, 093001 (2015).


\bibitem{Ablikim:2009aa} M.~Ablikim $et~al.$ [BESIII Collaboration], Nucl.\ Instrum.\ Meth.\ A {\bf 614}, 345 (2010).
\bibitem{geant4} S. Agostinelli $et~al.$ [GEANT4 Collaboration],
Nucl. Instrum. Meth. A {\bf 506}, 250 (2003).
\bibitem{kkmc} S. Jadach, B. F. L. Ward and Z. Was, Comput. Phys. Commun. {\bf 130}, 260 (2000); Phys. Rev. D {\bf 63}, 113009 (2001).
\bibitem{nima462_152} D. J. Lange, Nucl. Instrum. Meth. A {\bf 462}, 152 (2001); R. G. Ping, Chin. Phys. C {\bf 32}, 599 (2008).
\bibitem{SJNP41_466} E. A. Kurav and V. S. Fadin, Sov. J. Nucl. Phys. {\bf 41}, 466 (1985).
\bibitem{plb303_163} E. Richter-Was, Phys. Lett. B {\bf 303}, 163 (1993); E. Barberio and Z. Was, Comput. Phys. Commun. {\bf 79}, 291 (1994).
\bibitem{pdg2014} K. A. Olive $et~al.$ [Particle Data Group], Chin. Phys. C {\bf 38}, 090001 (2014) and 2015 update.


\bibitem{lundcharm} J. C. Chen, G. S. Huang, X. R. Qi, D. H. Zhang, Y. S. Zhu, Phys. Rev. D {\bf 62}, 034003 (2000).



\bibitem{1511.08380} M.~Ablikim $et~al.$ [BESIII Collaboration], Phys. Rev. Lett. {\bf 116}, 052001 (2016).
\bibitem{prd79_052010} J. Y. Ge $et~al.$ [CLEO Collaboration], Phys. Rev. D {\bf 79}, 052010 (2009).
\bibitem{Hinson:2004pj}  J.~W.~Hinson {\it et al.} [CLEO Collaboration], Phys. Rev. Lett. {\bf 94}, 191801 (2005).


\end{thebibliography}
\end{document}